\documentclass{sig-alternate}
\usepackage[english]{babel}
\usepackage[utf8]{inputenc}

\usepackage{lmodern}
\usepackage[T1]{fontenc}
\usepackage{xspace}
\usepackage{microtype}

\usepackage{tikz}
\usetikzlibrary{positioning,arrows,shapes}

\usepackage[pdftex, colorlinks=false]{hyperref}

\usepackage{booktabs}
\usepackage{rotating}
\usepackage{multirow}

\usepackage{amssymb}
\usepackage{amsmath}
\usepackage{stmaryrd}

\usepackage{wrapfig}

\usepackage[nolineno,norules]{lgrind}
\def\LGfsize{\scriptsize}
\def\CMfont{\ttfamily}
\def\KWfont{\ttfamily\bfseries}
\def\VRfont{\ttfamily}
\def\BGfont{\ttfamily}

\usepackage{listings}

\newcommand{\ie}{\emph{i.e.,}}
\newcommand{\eg}{\emph{e.g.,}}

\newcommand{\etal}{\emph{et al.}}
\newcommand{\archipattern}{SCC architectural pattern}
\newcommand{\contextop}{context operator}
\newcommand{\domainappl}{SCC application}

\newcommand{\myparagraph}[1]{\vspace{0.3\baselineskip}\noindent\emph{#1.}}

\newcommand{\margin}{
\setlength{\topsep}{0pt}
\setlength{\parskip}{0pt}
\setlength{\partopsep}{0pt}
\setlength{\parsep}{0pt}         
\setlength{\itemsep}{0pt} 
}

\newif\ifshowcomments
\showcommentstrue

\ifshowcomments
\newcommand{\mynote}[2]{\fbox{\bfseries\sffamily\scriptsize#1}
  {$\blacktriangleright$\textsf{\emph{#2}}$\blacktriangleleft$}}
\else
\newcommand{\mynote}[2]{}
\fi

\newif\iffinal
\finalfalse

\iffinal

\else

\fi

\newcommand{\msf}[1]{\mbox{\sf{{#1}}}}

\newcommand{\config}[3]{\ensuremath{\langle{#1};{#2};{#3}\rangle}}

\newcommand{\names}[1]{\ensuremath{\msf{names}(#1)}}

\newcommand{\denot}[1]{\ensuremath{\llbracket #1 \rrbracket}}
\newcommand{\typeof}[1]{\ensuremath{\mathit{typeof}(#1)}}
\newcommand{\accesstypeof}[1]{\ensuremath{\mathit{access\_typeof}(#1)}}
\newcommand{\publish}[1]{\ensuremath{\mathit{publish}(#1)}}
\newcommand{\args}[1]{\ensuremath{\mathit{args}(#1)}}
\newcommand{\code}[1]{\ensuremath{\mbox{\texttt{#1}}}}

\newcommand{\ct}[1]{{\normalsize\ttfamily #1}}

\newtheorem{definition}{Definition}

\begin{document}

\conferenceinfo{ICSE}{'11, May 21--28, 2011, Waikiki, Honolulu, HI, USA}
\CopyrightYear{2011}
\crdata{978-1-4503-0445-0/11/05}

\title{Leveraging Software Architectures to Guide and Verify the
  Development of Sense/Compute/Control Applications}


%
%
%
%

\numberofauthors{4} 
%
\author{
%
%
\begin{tabular}{cccc}
  Damien Cassou & Emilie Balland & Charles Consel & Julia Lawall \\
  \multicolumn{3}{c}{INRIA/University of Bordeaux} & DIKU/INRIA/LIP6 \\
  \multicolumn{3}{c}{{first.last}@inria.fr} & julia@diku.dk \\
\end{tabular}
}

\maketitle
\begin{abstract}
A software architecture describes the structure of a computing system by
specifying software components and their interactions. Mapping a software
architecture to an implementation is a well known challenge. A key
element of this mapping is the architecture's description of the data and
control-flow interactions between components. The characterization of these
interactions can be rather abstract or very concrete, providing more or
less implementation guidance, programming support, and static verification.

\looseness -1
In this paper, we explore one point in the design space between
abstract and concrete component interaction specifications. We
introduce a notion of \emph{interaction contract} that expresses
allowed interactions between components, describing both data and
control-flow constraints. This declaration is part of the architecture
description, allows generation of extensive programming support, and
enables various verifications. We instantiate our approach in an
architecture description language for Sense/Compute/Control
applications, and describe associated compilation and verification
strategies.


\end{abstract}




\section{Introduction}
\label{sec:introduction}

A Sense/Compute/Control (SCC) application is one that interacts with the
physical environment~\cite{Tayl09a}.  Such applications are
pervasive in domains such as building automation, assisted living, and autonomic
computing. Developing an \domainappl{} is complex because the implementation
must address both the interaction with the environment and the application
logic, because any evolution in the environment must be reflected in the
implementation of the application, and because correctness is essential, as
effects on the physical environment can have irreversible consequences.

We have observed that SCC applications can be defined according to an
architectural pattern involving four kinds of components, organized
into layers~\cite{Edwar09a}: (1) \emph{sensors} at the bottom, which
obtain information about the environment; (2) then \emph{context
  operators}, which process this information; (3) then \emph{control
  operators}, which use this refined information to control (4)
\emph{actuators} at the top, which finally impact the environment.
Data and control-flow interactions between these layers are restricted.
Sensors may be proactive and initiate data flows when they detect
changes in the environment, while the other kinds of components are
only reactive. Context operators may receive information from sensors
or other context operators, and may interrogate the same, to obtain
further information. Control operators can receive information only
from context operators and actuators are only activated by orders,
such as ``turn on'' or ``send this email'', sent by the control
operator layer.  While our experience shows that this SCC
architectural pattern captures the architecture of many kinds of SCC
applications, the question remains of how to exploit it to guide an
implementation.

When a software architecture is expressed formally using an Architecture
Description Language (ADL)~\cite{Medv00a}, and is sufficiently concrete, it
may be possible to generate an implementation automatically. But this
requires providing a complete description of the application behavior in
the architecture, which mixes concerns, obscures the interaction
constraints, and defeats reusability. In particular, the SCC architectural
pattern describes application design at a more abstract level, in that it
does not incorporate the application logic. Mapping such an abstract
software architecture into an implementation and maintaining the
relationship between the architecture and the implementation as they evolve
are well known to be complex tasks~\cite{Tayl09a}.

Several recent approaches have considered the relation between
architecture and implementation, focusing on the interaction between
components. One such approach is ArchJava that embeds an ADL into a
programming language to allow architectural concerns to be part of the
application code~\cite{Aldr02c}. This approach however, entails a
mixing of the architecture and implementation that may obscure both of
them. To regain separation of concerns between architecture
description and implementation, Archface proposes a new interface
mechanism~\cite{Ubay10a}, leveraging concepts from Aspect-Oriented
Programming (AOP) to describe component interactions. AOP pointcuts
abstract the structure of implementations, providing constraints such
as when a particular method must be called during a given control
flow. Such approaches make it possible to verify that an
implementation conforms to an architecture description. Still, both of
these approaches blur the separation between architecture description
and implementation, making the architectural design phase more
difficult.

In this paper, we propose an approach to linking architecture and
implementation that specifically targets \domainappl{}s. Our approach
balances the abstraction and concreteness of the architecture description
by introducing a notion of \emph{interaction contract}. An interaction
contract declares what interactions a given component can perform,
expressing in high-level terms both data and control-flow constraints. This
declaration is part of the architecture description, keeping this phase
separated from the implementation.  Yet, our interaction contracts allow
the architect to precisely specify the interactions between components,
without simultaneously having to reason about code structure. Interaction
contracts furthermore can be used to generate extensive programming
support, ensuring the conformance between the architecture and the
implementation and guiding the development phase. The architect can also use
the constraints expressed by interaction contracts to verify a range of
properties beyond implementation conformance.

\paragraph*{Contributions}
In this paper, we introduce an architecture-driven generative
methodology that improves the design, programming and verification of
\domainappl{}s. Our contributions are as follows.

\begin{itemize}
\margin
\item We introduce a language for interaction contracts dedicated to
  SCC applications (Section~\ref{sec:pattern}).
\item We show that interaction contracts can guide the implementation
  of \domainappl{}s by enabling the generation of highly customized
  programming frameworks using a dedicated compiler
  (Section~\ref{sec:generation}). This approach ensures that
    the architecture conforms to the implementation, while
    facilitating software evolution.
\item We show that such interaction contracts are precise enough to verify
  safety properties such as information flow reachability or
  interaction invariants (Section~\ref{sec:analysis}).
\item We extend our previously developed implementation of an ADL
  targeting SCC applications~\cite{Cass09b} with interaction contracts,
  and use this implementation to assess the benefit of interaction
  contracts at a conceptual level and in terms of metrics on the
  resulting code (Section~\ref{sec:discussions}).
\end{itemize}

\section{Our Approach}
\label{sec:pattern}

We first present the SCC architectural pattern~\cite{Edwar09a} and then
introduce the notion of interaction contract. The \archipattern{} is
based on the \emph{sense/compute/control} pattern presented by Taylor
\etal{}~\cite{Tayl09a} and on the pattern presented by Chen
and Klotz~\cite{Chen02b} for ubiquitous computing systems. Interaction
contracts enrich the \archipattern{} to describe interactions among
components.


\subsection{SCC Architectural Pattern}
\label{subsec:pattern}

We first introduce the terminology and concepts used throughout the
paper. The data flow of the SCC architectural pattern is expressed by
an oriented graph whose nodes are the architecture components and
whose (solid) edges indicate data exchange between components (see
Figure~\ref{fig:webserver-simple}). We say that the \emph{children}
and \emph{parents} of a component are respectively the sources of the
incoming edges and targets of the outgoing edges connected to the
component. There are two types of interactions a component can
perform: \emph{pushing} data to the parents or responding to a
\emph{pull} request from one of its parents. Pull requests
  are represented in the graph as dashed edges, and can be
  parameterized.

The \domainappl{} pattern involves four layers: sensors, context
operators, control operators, and actuators. Each layer corresponds to
a separate class of components:
\begin{itemize}
\margin
\item \emph{Sensors} send information sensed from the environment to the
  context operator layer through data \emph{sources}.  Sensors can both
  push data to context operators and respond to context operator requests.
  We use the term ``sensor'' both for entities that actively
  retrieve information from the environment, such as system probes, and
  entities that store information previously collected from the
  environment, such as databases.

\item \emph{Context operators} refine (aggregate and interpret) the
  information given by the sensors. 
  Context operators can push data to other context operators and to
  control operators. Context operators can also respond to requests
  from parent context operators.
\item \emph{Control operators} transform the information given by the
  context operators into orders for the actuators. 
\item \emph{Actuators} trigger actions on the environment. 
\end{itemize}

Sensors are \emph{proactive} or \emph{reactive} components whereas
context operators, control operators and actuators are always
\emph{reactive}. These properties ensure that \domainappl{}s are
reactive to the environment state. That is, all computation
is initiated by a publish/subscribe interaction with a sensor.

As the underlying architecture is component-based, the application can
be fully distributed. To prevent concurrent handling of events in a
component, all interactions of a component are queued and executed one
at a time, sequentially.

\subsection{Example}
\label{subsec:example}

\begin{figure}[htbp]
\centering
\includegraphics[width=0.9\linewidth]{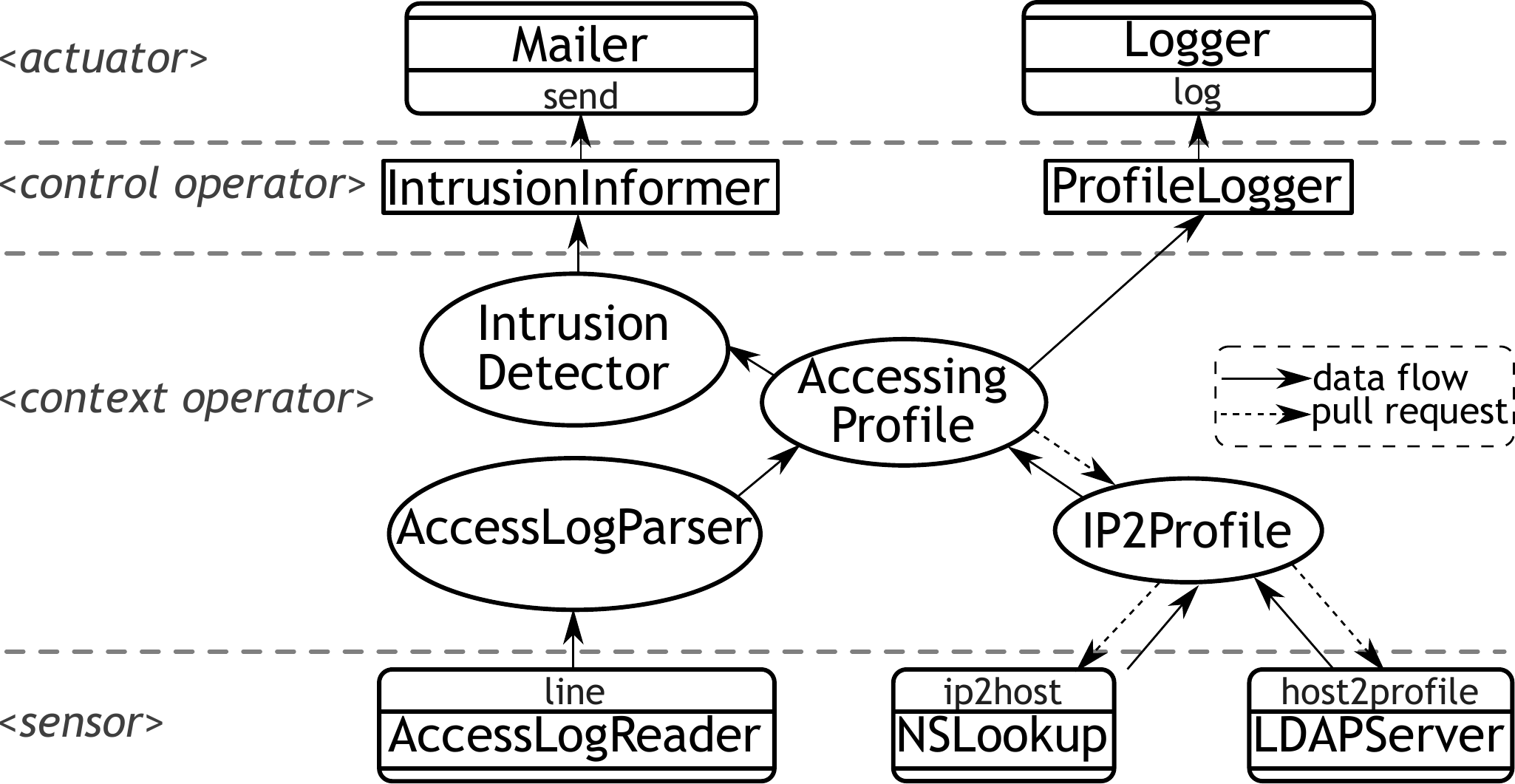}
\caption{Architecture of a web server monitor. Solid arrows represent
data flow. Dashed arrows represent pull requests. For
simplicity, the diagram does not show the types of the values
calculated by the components and the types of the parameters required
by pull requests.
\label{fig:webserver-simple}} 
\end{figure}

As a running example, we define the architecture of a web server
monitoring application. In this SCC application, the considered
environment is a tier system, consisting of a web server and
associated network tools. The two tasks that we want to implement are
(1) updating a log containing profiles of the web server's clients
(client name and IP address) and (2) sending an email to
administrators in case of intrusions.
Figure~\ref{fig:webserver-simple} illustrates the component layers and
the data-flow interactions between the components of this
application.

\myparagraph{Sensors} The state of the environment is observed using
three sensors. 1) The access log reader provides one information
source named \texttt{line}. This information source is updated when a new line is
added to the log for an access to the server. 2) The NSLookup tool
returns the host name associated with an IP address via the
information source named \texttt{ip2host}. 3) The LDAP server returns
the profile of a host name via the information source named
\texttt{host2profile}.

\myparagraph{Context operators for profile identification} The
\texttt{Ac\-cess\-ing\-Profile} context operator, in the middle of
Figure~\ref{fig:webserver-simple}, calculates which profile is
accessing the web server. This context operator is activated by the
\texttt{AccessLogParser} context operator, which is itself activated
by the information source \texttt{line} of the
\texttt{Access\-Log\-Reader} sensor. When a new line is added to the
log, \texttt{Ac\-cess\-Log\-Parser} parses the line to create a
higher-level structure including the IP address of the person
accessing the web server, and the requested page. This information is
passed to \texttt{AccessingProfile}, which extracts the IP address
from the structure, and then asks the \texttt{IP2Profile} context
operator to compute a profile. This profile is obtained by querying
the \texttt{NSLookup} and the \texttt{LDAPServer} sensors. Pull
requests on \texttt{IP2Profile} and \texttt{NSLookup} are
parameterized by an IP address. Pull requests on \texttt{LDAPServer}
are parameterized by a host name.

\myparagraph{Context operators for intrusion detection} The
\texttt{In\-tru\-sion\-Detector} context operator uses the context
calculated by \texttt{Ac\-ces\-sing\-Profile}, including the
information about the most recent access to the web server and the
client profile associated with this access.
\texttt{In\-tru\-sion\-Detector} only propagates accesses that are
suspected to represent intruders.

\myparagraph{Control operators and actuators} The monitoring tasks are
implemented by the \texttt{Intrusion\-Informer} and
  \texttt{ProfileLogger} control operators, which respectively invoke
the \texttt{Mailer} and \texttt{Logger} actuators using the
\texttt{send} and \texttt{log} actions. To notify the administrator of
an intrusion, \texttt{Intrusion\-Informer} only needs to be informed
by the \texttt{Intrusion\-Detector} context operator of any new
intrusion. To update the profile log, \texttt{ProfileLogger} only
needs to be informed by the \texttt{AccessingProfile} context operator
of which profile is accessing the server.

\medskip

In this example, we can observe that the architecture description in
Figure~\ref{fig:webserver-simple} is underspecified. While it may be
intuitively obvious that the \texttt{IP2Profile} context operator
reacts only to parent pull requests, as \texttt{LDAPServer} and
\texttt{NSLookup} never push data by themselves, this information is
not explicit in the architecture description. This underspecification
may lead to different interpretations of the architecture description
and incompatible implementations. To address these issues, we enrich
the architecture description by annotating each context operator with
an \emph{interaction contract}.

\subsection{Interaction Contracts}
\label{subsec:behavcontract}

The goal of an interaction contract is to describe the interactions
that are allowed by the context operators of an SCC application. In a
reactive system, the most basic information is what makes a context
operator react, \ie{} a data pull request from one of its parents or a
data push from its children. In this reaction, a context operator may
need to pull data from its child context operators or child sensor
sources. Finally, a reaction may or may not lead to the push of a new
value. We group the information about these three kinds of
interactions into a \textit{basic interaction contract}.

\begin{definition}
  A basic interaction contract $\config{A}{U}{E}$ is a tuple where $A$,
  $U$ and $E$ are named respectively the activation condition, the data
  requirements list and the emission. These elements are defined as
follows:
\begin{itemize}
\margin
\item $A = {\Uparrow(A_1,\ldots,A_n)} \mid {\Downarrow
    \mathit{self}}$, where $n>0$, $A_i$ is the name of a child of the
  current context operator (a sensor source or a context operator) or
  a disjunction of such names, and self indicates the context
  operator itself. ${\Uparrow(A_1,\ldots,A_n)}$ corresponds to the
  push of values from all the children $A_1,\ldots,A_n$. If any $A_i$
  is a disjunction of names, then the information associated with any
  of these names can be used. $\Downarrow \mathit{self}$ corresponds
  to a pull request from a parent of the context operator. A pull
  request always returns a value to the calling parent.
\item $U = {\Downarrow(B_1,\ldots,B_n)}$ where $n\geq 0$ and $B_i$ is
  the name of a child of the current context operator (a sensor source
  or a context operator). This information is accessed by a pull
  request, and the developer may choose to access it or not.
\item $E = {\Uparrow \mathit{self}} \mid {\Uparrow \mathit{self}?}
  \mid \emptyset$ indicates respectively whether the context operator
  always, sometimes, or never pushes a new value to all its parents.
  When $A = {\Downarrow \mathit{self}}$, a value is always returned to
  the requesting parent, regardless of $E$.
\end{itemize}

\end{definition}

\noindent
An interaction contract defines how a context operator interacts with
its parents and children, and in this sense is related to interaction
descriptions such as automata-based models~\cite{Alfaro01,Lynch87}, as
analyzed in Section~\ref{sec:related}.

Table~\ref{tab:interaction-contracts} specifies the interaction contracts for
the web server
monitoring architecture. For example, the interaction contract
of the \texttt{Intrusion\-Detector} indicates, via the notation ${\Uparrow
  \mathit{self?}}$, that when \texttt{Intrusion\-Detector} receives a new
profile from \texttt{Accessing\-Profile}, it might or might not push a
profile. In practice, \texttt{Intrusion\-Detector} only pushes a profile
when the profile is suspected to correspond to an intrusion.  In contrast,
the emission of the interaction contract associated with \texttt{IP2Profile}
is $\emptyset$. When this component receives a pull request, it returns
the data to the parent that sent the request, but it does not inform the
other parents, if any, by publishing the data.

\begin{table}[tbp]
  \centering
  \scriptsize
  \begin{tabular}[htbp]{@{}ll@{}}
\toprule
\textbf{Context operator} & \textbf{Associated interaction contract}\\
\midrule
AccessLogParser & $\config{{\Uparrow (\code{line})}}{\emptyset}{{\Uparrow \mathit{self}}}$\\
AccessingProfile & $\config{{\Uparrow(\code{AccessLogParser})}}{{\Downarrow (\code{IP2Profile})}}{{\Uparrow \mathit{self}}}$\\
IP2Profile & $\config{{\Downarrow \mathit{self}}}{{\Downarrow(\code{ip2host},\code{host2profile})}}{\emptyset}$\\
IntrusionDetector & $\config{{\Uparrow(\code{AccessingProfile})}}{\emptyset}{{\Uparrow \mathit{self}?}}$\\
\bottomrule
  \end{tabular}
  \caption{Interaction contracts associated to the context operators of the web server
    architecture. \texttt{Line}, \texttt{ip\-2\-host} and
    \texttt{host\-2\-profile} abbreviate \texttt{Access\-Log\-Reader.line},
    \texttt{NS\-Look\-up.ip\-2\-host} and
    \texttt{LDAP\-Server.host\-2\-profile}, respectively.}
  \label{tab:interaction-contracts}
\end{table}

\begin{wrapfigure}{L}{3.5cm}
\centering
\includegraphics[width=1\linewidth]{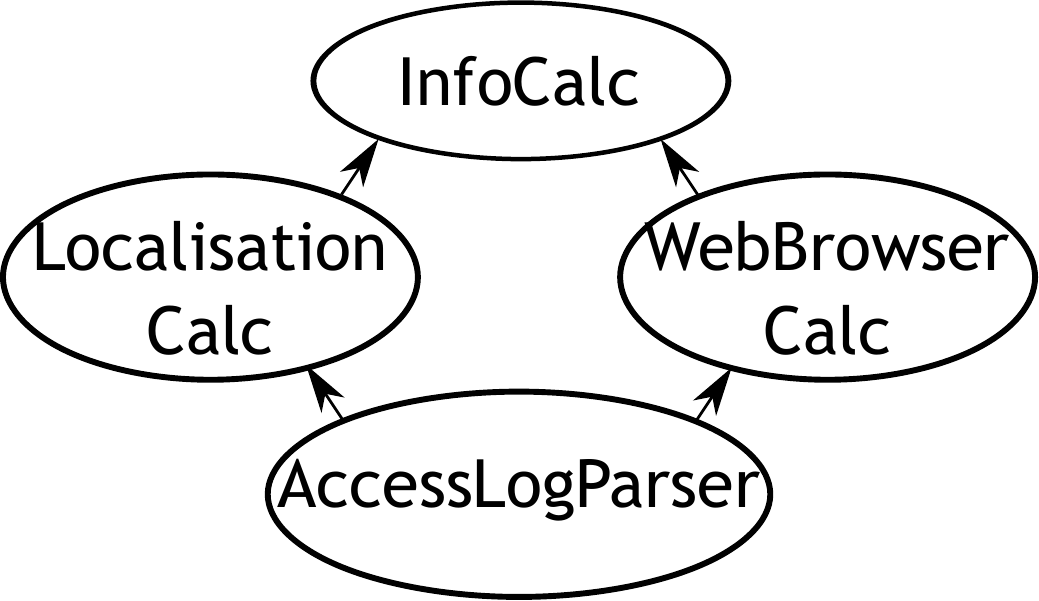}
 \caption{Example of synchronization.
\label{fig:webserver-statistic}}
\end{wrapfigure}

\myparagraph{Synchronization} A sequence $\Uparrow(A_1,\ldots,A_n)$ in
the activation condition of an interaction contract indicates the
synchronization of multiple information sources. Suppose that the
context calculated by the \texttt{Access\-Log\-Par\-ser} context
operator were refined into two types of contexts: the geographic
location of the host and the web browser used for this access,
represented by the context operators
\texttt{Lo\-ca\-li\-za\-tion\-Calc} and \texttt{Web\-Brow\-ser\-Calc},
respectively (see Figure~\ref{fig:webserver-statistic}). The
information calculated by these context operators is then combined
using a new context operator \texttt{InfoCalc}. Its interaction
contract is
$\config{{\Uparrow(\code{Web\-Brow\-ser\-Calc},\code{Lo\-ca\-li\-zationCalc})}}{\emptyset}{{\Uparrow
    \mathit{self}}}$, which ensures that we obtain synchronised
information from \texttt{LocalizationCalc} and
\texttt{WebBrowserCalc}.

\myparagraph{Disjunction} A disjunction of names in an activation
condition indicates that a context operator can use any one of
multiple distinct contexts. For example, in a web server, a dangerous
access can be due either to an intrusion or to an SQL injection. The
information about both conditions has the same type, \texttt{Access}.
Suppose we have a new context operator \code{SQLInjDetector} that
pushes \texttt{Access} data when there is an SQL injection. Then, we
can define a \texttt{DangerDetection} context operator that
abstracts over these two types of danger using the interaction contract:
$\config{{\Uparrow(\code{IntrusionDetector} \vee
    \code{SQLInjDetector})}}{\emptyset}{{\Uparrow \mathit{self}}}$

\myparagraph{Interaction contract composition} As a context operator can be
activated by different conditions, possibly leading to different behaviors,
we introduce the $\parallel$ operator that allows the combination of
several basic interaction contracts.  A composition of basic interaction
contracts is, for example, necessary when a context operator can be activated
by both a data pull from one of its parents and a data push from one of its
children. For example, the \texttt{AccessingProfile} context operator could
have a second interaction contract $\config{\Downarrow
  \mathit{self}}{\emptyset}{\emptyset}$ that allows access to the most
recent value of this context at any moment in the execution of the
application.


\subsection{Architecture Consistency and Determinacy}

An interaction contract of a context operator implies interaction
requirements on the operator's parents and children. For example,
given a context operator \texttt{A}, the existence of some interaction contract
whose data
requirement is $\Downarrow{} \mbox{\texttt{A}}$ implies that \texttt{A} has an
interaction contract whose activation condition is $\Downarrow{}
\mathit{self}$. An architecture is \emph{consistent} when each of its
interaction contracts respects the requirements imposed by all other
interaction contracts of the architecture.

\looseness -1 Furthermore, a given data flow should not trigger the
activation of multiple basic interaction contracts of a single context
operator.  This situation occurs, for example, if the activation
conditions of two different interaction contracts of a single context
operator are both pull requests. An architecture is
\emph{deterministic} if no context operator has a pair of basic
interaction contracts that are activated by the same data flow.

Given a basic interaction contract $\config{A}{U}{E}$, let $\names{A}$
be the set of names of sensor sources or context operators
(\textit{self} included) used in $A$. For example,
$\names{\Uparrow(P \vee Q,R)}=\{P,Q,R\}$ and $\names{\Downarrow
  \mathit{self}}=\{\mathit{self}\}$.


\begin{definition}[Contract Consistency]
Given an architecture $\Delta$, an interaction contract $\alpha$ is
consistent relative to $\Delta$ if one of the following conditions is
satisfied:

\begin{itemize}
\margin
\item if $\alpha = \config{A}{\Downarrow(B_1,\ldots,B_m)}{E}$ then:
\begin{itemize}
\margin
\item for each $B_i$ that is a context operator, there is a
behavioral contract $\config{\Downarrow \mathit{self}}{\_}{\_}$ associated with
$B_i$ in $\Delta$,
\item if $A = {\Uparrow(\ldots)}$ then for each $N \in \names{A}$
that is a context operator, there is an interaction contract
$\config{\_}{\_}{{\Uparrow \mathit{self}}}$or
$\config{\_}{\_}{{\Uparrow \mathit{self} ?}}$ associated with $N$ in
$\Delta$.
\end{itemize} 
\item if $\alpha = \alpha_1 \parallel \ldots \parallel \alpha_n$ then
each $\alpha_i$ is consistent relative to $\Delta$.
\end{itemize} 
\end{definition}

\begin{definition}[Architecture Consistency]
An architecture $\Delta$ is consistent if each interaction contract
associated with its context operators is consistent relative
to~$\Delta$.
\end{definition}

\begin{definition}[Contract Interference]
\label{definition:interference}
A basic interaction contract $\config{A}{U}{E}$ interferes with a
basic interaction contract $\config{A'}{U'}{E'}$ if $\names{A} \cap
\names{A'} \neq \emptyset$.
\end{definition}

\begin{definition}[Contract Determinacy]
\label{definition:determinicity}
An interaction contract $\alpha_1 \parallel \dots \parallel \alpha_n$ is
deterministic if each basic interaction contract $\alpha_i$ does not
interfere with any of the others.
\end{definition}

\begin{definition}[Architecture Determinacy]
An architecture is deterministic if all its interaction contracts are
deterministic.
\end{definition}

To ensure consistency and determinacy of an architecture, the
architecture compiler should enforce these properties. 

\subsection{Interaction Contract Semantics}
\label{sec:sem}

A context operator can be viewed as a function, as it reacts to some
inputs and potentially produces an output. Thus, the denotational
semantics of an interaction contract is a
function type. Intuitively, each possible implementation of a
  context operator, whose interaction contract is $C$, corresponds to
  a function of type \denot{C}, as defined in
  Table~\ref{table:denotation}. In this table, $O$ represents a
  context operator and $T$ its return type. Essentially, the
  activation condition and the data requirements determine the types of the
  parameters of this function, and the emission determines its return
  type. Each data requirement maps to the type of a callback function
  that takes the request parameters as arguments. If a context
  operator is associated with several interaction contracts (denoted
  $C_1 \parallel \ldots \parallel C_n$), then the type of the context
  operator is represented as a tuple of functions $\denot{C_1} \times
  \ldots \times \denot{C_n}$.

\newcommand{\piify}[2]{\Pi_{i=1}^{#2}#1}

\begin{table*}[!hbtp]
  \centering
      \begin{tabular}{l@{~=~}l}

$\denot{\config{{\Uparrow(A_1,\ldots,A_n)}}{{\Downarrow(B_1,\ldots,B_m)}}{{\Uparrow \mathit{self}}}}$
& $\piify{\typeof{A_i}}{n} \times \piify{\accesstypeof{B_i}}{m} \rightarrow T$ \\

$\denot{\config{{\Uparrow(A_1,\ldots,A_n)}}{{\Downarrow(B_1,\ldots,B_m)}}{{\Uparrow \mathit{self}?}}}$
& $\piify{\typeof{A_i}}{n} \times \piify{\accesstypeof{B_i}}{m} \times \publish{T} \rightarrow ()$\\

$\denot{\config{{\Uparrow(A_1,\ldots,A_n)}}{{\Downarrow(B_1,\ldots,B_m)}}{\emptyset}}$
& $\piify{\typeof{A_i}}{n} \times \piify{\accesstypeof{B_i}}{m} \rightarrow ()$\\

$\denot{\config{\Downarrow {\mathit{self}}}{\Downarrow(B_1,\ldots,B_m)}{E}}$
& $\args{O} \times \piify{\accesstypeof{B_i}}{m} \rightarrow T$ if $E={\Uparrow \mathit{self}}$ or $E=\emptyset$\\

$\denot{\config{\Downarrow {\mathit{self}}}{\Downarrow(B_1,\ldots,B_m)}{\Uparrow \mathit{self}?}}$
& $\args{O} \times \piify{\accesstypeof{B_i}}{m} \times \publish{T} \rightarrow T$\\

$\denot{C_1\!\parallel \ldots \parallel\! C_n}$
& $\denot{C_1} \times \ldots \times \denot{C_n}$\\

  \end{tabular}


    \caption{Denotation of an interaction contract.}
    \label{table:denotation}
\end{table*}

The rules in Table \ref{table:denotation} use the functions
\textit{args}, \textit{publish}, \textit{typeof}, and
\textit{access\_typeof}, defined as follows. Given an identifier $n$
that is the name of a context operator or a sensor source, the
function $\typeof{n}$ maps $n$ to its type. For example, in the web
server application:

\vspace{.3em}

\begin{tabular}{l@{~=~}l}
$\typeof{\code{line}}$ & $\code{String}$ \\
$\typeof{\code{AccessLogParser}}$ & $\code{Access}$ \\
$\typeof{\code{AccessingProfile}}$ & $\code{Profile}$ \\    
\end{tabular}

\vspace{.7em}

\noindent
Note that $\typeof{A_1 \vee A_2}= \typeof{A_1}\;\cup\;\typeof{A_2}$
where $\cup$ denotes the operator for the union type (in Java, for
example, $t_1 \cup t_2$ is the smallest common supertype of $t_1$ and
$t_2$, which is at worst the type \texttt{Object}). The function
\accesstypeof{n} maps $n$ to the type of a data pull request:
$$\accesstypeof{n} = \args{n} \rightarrow \typeof{n}$$ where
$\args{n}$ returns the types of the parameters needed for a pull
request on $n$. For example,

\vspace{.3em}

\begin{tabular}{l@{~=~}l}
$\accesstypeof{\code{ip2host}}$ &$\code{IPAddress} \rightarrow \code{String}$ \\
$\accesstypeof{\code{host2profile}}$ & $\code{String} \rightarrow \code{Profile}$\\
\end{tabular}

\vspace{.7em}

\noindent
Finally, given a type $T$, we denote by $\publish{T}$ the type $T
\rightarrow ()$, which is the type of a function that publishes data of
type $T$. 

In the web server example, the denotation of the interaction contract
$\config{{\Uparrow(\code{AccessLogParser})}}{{\Downarrow
(\code{IP2Profile})}}{{\Uparrow \mathit{self}}}$ associated to
\texttt{Accessing\-Profile} is the function type $\code{Access} \times
(\code{IPAddress} \rightarrow \code{Profile}) \rightarrow
\code{Profile}$.

\subsection{Design Support}

The architectural pattern presented here is to be used as a paradigm
guiding the architect in the decomposition of an \domainappl{} into
layers of components. The interaction contracts enable an architect to
describe the allowed interactions among components. Interaction contracts
are limited to the kinds of interactions possible within the SCC
architectural pattern, further guiding the architect.


\section{Programming Support}
\label{sec:generation}

We have integrated interaction contracts into DiaSpec, our
domain-specific ADL for SCC applications~\cite{Cass09b}. From a
Dia\-Spec description, a compiler produces a dedicated Java
programming framework that is both \emph{prescriptive} and
\emph{restrictive}: it is prescriptive in the sense that it guides the
developer, and it is restrictive in the sense that it limits the
developer to what the architecture allows. In this section, we
describe the compilation strategies that achieve this.

\subsection{Structure of the Generated Code}
\label{sec:class_structure}

Our compiler takes as input an architecture description written in the
DiaSpec ADL. This specification describes textually an instance of the
\archipattern{} and the associated interaction contracts. From this
architecture description, the compiler generates a dedicated
programming framework containing support for sensors, context
operators, control operators, and actuators. We focus on the code
generated for interaction contracts as the other parts of the
framework have been described previously~\cite{Cass09b}.

For each context operator declared in the architecture description,
the compiler generates an abstract class. The abstract methods in this
class represent code to be provided by the developer, to allow him to
program the application logic (\eg{} to answer a pull request). To
implement the context operator, the developer implements these methods
in a subclass of this abstract class.

For each basic interaction contract, the generated abstract class
contains an \emph{abstract method} and a corresponding \emph{calling
  method}. The abstract method is to be implemented by the developer
while the calling method is used by the framework to call the
implementation of the abstract method with the expected arguments. The
translation of each interaction contract of the architecture into an
abstract method is similar to the denotation found in
Table~\ref{table:denotation}. Callbacks are used to encapsulate an
optional interaction between the context operator and its parents or
one of its children. The developer may decide to invoke each callback
depending on his needs. A callback is only provided when an
interaction is optional. If the interaction is mandatory, it is
automatically done by the calling method. If the interaction is
forbidden, the calling method does not provide the developer with any
means to perform the interaction.


\subsection{Interaction Contract Compilation}

To illustrate the compilation process, we consider the Java code
generated for some of the interaction contracts of the web server
monitoring architecture (see Table~\ref{tab:interaction-contracts}).

The \texttt{AccessingProfile} interaction contract is
compiled into the following abstract method:

{\scriptsize
\begin{verbatim}
abstract IdentifiedAccess onNewAccessLogParser(Access newAccess,
                         PullFromIP2ProfileCallback ip2Profile);
\end{verbatim}
}

\noindent
The name of this abstract method starts with \texttt{onNew}, reflecting
the fact that the child is providing a new value. The method's first
parameter represents the new parsed line and the second parameter
represents a callback function that
permits a pull interaction with \ct{IP2Profile}. This callback takes an
\ct{IPAddress} as argument and returns a corresponding \ct{Profile}.
The return type of the \texttt{onNewAccessLogParser} abstract method
forces the implementation of the abstract method to return a profile, which
is pushed automatically by the calling method upon method return.

\ct{IP2Profile} is a kind of database that can only be accessed
through pull requests with an \ct{IPAddress} as an argument. From its
interaction contract, the following abstract method is generated:

{\scriptsize
\begin{verbatim}
abstract Profile get(IPAddress newIPAddress,
                        PullFromNSLookupCallback ip2Host,
                        PullFromLDAPServerCallback host2Profile);
\end{verbatim}}

\noindent
The name of this abstract method is \texttt{get}, reflecting the fact
that the parent requested a value. The implementation of this abstract
method may call two sources through corresponding callbacks. Because
the emission is ${\emptyset}$, \ct{IP2Profile} returns a result only
to its requesting parent, \ie{} \texttt{IP2Profile} does not push its
value and has no way to do so.

From the interaction contract of \texttt{IntrusionDetector}, the following
abstract method is generated:

{\scriptsize
\begin{verbatim}
abstract void onNewAccessingProfile(
                          IdentifiedAccess newIdentifiedAccess,
                          PublishCallback publish);
\end{verbatim}
}

\noindent
Not all identified accesses arriving at \ct{IntrusionDetector} are
necessarily intrusions. The publish callback allows the application
logic in the method implementation to decide whether to give an alert about
an intrusion. 

\myparagraph{Synchronization} From the interaction contract of 
\texttt{InfoCalc}, \\
$\config{{\Uparrow(\code{WebBrowserCalc},\code{LocalizationCalc})}}{\emptyset}{{\Uparrow
    \mathit{self}}}$,
the following abstract method is generated:

{\scriptsize
\begin{verbatim}
abstract Info onNewWebBrowserCalcAndLocalizationCalc(
                         WebBrowser newWebBrowser,
                         Localization newLocalization);
\end{verbatim}
}

\noindent
This method is to be called with both values from the children as soon
as they are both present. Various strategies can be used to implement
this kind of synchronization: one approach is to remember only the
most recent value from each source, while another is to enqueue all
values. Our default implementation uses a queue for each source. When
each queue has at least one value, the framework consumes one value
from each queue and invokes the abstract method on the resulting tuple
of values. This implementation may be changed by the developer.

\myparagraph{Disjunction} From the interaction contract of
\texttt{Danger\-De\-tec\-tion},
$\config{{\Uparrow(\code{SQLInjDetector} \vee
    \code{IntrusionDetector})}}{\emptyset}{{\Uparrow \mathit{self}}}$,
the following abstract method is generated:

{\scriptsize
\begin{verbatim}
abstract IdentifiedAccess onNewDisjunction(
                         IdentifiedAccess newIdentifiedAccess);
\end{verbatim}
}

\noindent
This method has just one parameter to represent the disjunction. The
generated framework calls this method each time data is sent from
either \ct{SQLInj\-Detector} or \ct{In\-tru\-sion\-Detector}.

\myparagraph{Callbacks} As noted above, we use callbacks to implement
optional interactions with parents and children. Each callback is
implemented as an internal Java class, in the abstract class of the
context operator, and contains a single method. For example,
\texttt{Pull\-From\-IP2Profile\-Callback} is defined as:

{\scriptsize
\begin{verbatim}
public abstract class AbstractAccessingProfile {
  ...
  protected class PullFromIP2ProfileCallback {
    ...
    public Profile get(IPAddress ipAddress) {
        // pull the value from the instance of IP2Profile
        // through a call to the underlying middleware
        return ...;
    }
  }
}
\end{verbatim}
}

Callbacks are instantiated by the calling method, which passes them to
the abstract method. To ensure that the declared interaction contracts
are respected, we have to ensure that a callback is not invoked after
executing the code implementing the abstract method. \emph{I.e.,} the
developer must not store the callback for later use. Currently, this
property is not enforced statically, as it would
require adapting a Java compiler. Instead, we provide a dynamic guard
to prevent this situation at runtime. This dynamic guard is
implemented as a private \ct{boolean} variable in the internal class
whose value is checked before executing the callback. The dynamic
guard could also be extended to prevent the developer from calling a
callback more than a given number of times, \eg{} to limit resource usage.

\subsection{Programming Support}

The generated programming framework guides the developer with respect
to the architecture description. Implementing a declared component is
done by \emph{subclassing} the corresponding generated abstract class.
In doing so, the developer is forced to implement each abstract
method. To facilitate this process, most IDEs (such as Eclipse) for
object-oriented languages generate class templates based on abstract
super-classes. The generated framework passes each required piece of
information as an argument to the method. These arguments free the
developer from having to guess method or class names. The following
code presents a partial implementation as could be written by the
developer of the \ct{AccessLogParser} context operator.

{\scriptsize
\begin{verbatim}
 1: public class LighttpdAccessLogParser
                    extends AbstractAccessLogParser {
 2:   @Override
 3:   public Access onNewLine(String newLine) {
 4:     Access access = new Access();
 5:     access.setLine(newLine);
 6:     access.setHost_ip(parseRemoteHostIP(newLine));
 7:     ... // parsing of the other fields
 8:     return access;
 9:   }

10:   private IPAddress parseRemoteHostIP(String newLine) {
11:     Pattern pattern = Pattern.compile("^([^ ]+) ");
12:     Matcher m = pattern.matcher(newLine);
13:     m.find();
14:     return new IPAddress(m.group(1));
15:   }
16: }
\end{verbatim}
}

The method \texttt{onNewLine} is automatically called by the
programming framework when a new line arrives from an instance of
\texttt{AccessLogReader}. This method is typical of what has to be
implemented by the developer. Because most of the interaction details
are abstracted away by the generated framework, the developer can
concentrate on the application logic. For example, the decision of
whether or not to publish an \texttt{Access} value, and which
components are interested in this value, is abstracted away from the
implementation: on line~8, the developer simply returns the new value
without having to know what will happen to it.

\subsection{Ensuring Conformance}

An implementation must conform to its architecture. There are three
basic conformance criteria: decomposition, interface conformance and
communication integrity~\cite{Luck95a}.

\myparagraph{Decomposition} ``For each component in the architecture, there
should be a corresponding component in the implementation.''  This property
is satisfied in the sense that at least an abstract class is generated for
each component; nevertheless, the framework is not able to force the
developer to implement the full set of abstract classes.

\myparagraph{Interface Conformance} ``Each component in the
implementation must conform to its architectural interface.'' Our
compiler generates an abstract class that conforms by construction to
the component description. By extending the abstract class, the
component implementation automatically also conforms to this description.

\myparagraph{Communication Integrity} ``Each component in the
implementation may only communicate directly with the components to which
it is connected in the architecture.'' This property is satisfied because:
(1) an interaction only happens during the execution of an \ct{onNew} or
\ct{get} method, and only through the provided callbacks. (2) A component
never gains a direct reference to another component, and thus it can never
give such a reference to another component.

\subsection{Support for Evolution}

Maintenance and evolution are important parts of the development of
any software system. Our code generation strategy limits the number of
code changes required when the architecture description changes. When
this happens, the framework can be regenerated without overwriting the
developer's implementation. Any mismatches between the existing code
and the new programming framework are revealed by the Java compiler.
This strategy contrasts with strategies based on generating source
code skeletons to be filled by the developer, which mix
manually-written and generated code. In many of these strategies,
regenerating a skeleton overwrites the developer's implementation.


\section{Verification Support}
\label{sec:analysis}

In \domainappl{}s, safety is a key requirement as unexpected behaviors
can directly impact the environment and users through actuators.
Interaction contracts make explicit valuable information about the
data flow in the design and allow design-time safety verifications.
For example, with interaction contracts, it is possible to know at
design time all the context operators that will eventually be
activated by the publication of a given source. Moreover, our
generative approach ensures that these properties will be preserved at
the implementation level. To illustrate the possible design-time
analyses, we consider two kinds of properties:
\begin{itemize}
\margin
\item \emph{Data reachability}: can a component unexpectedly access
  critical data from a sensor or a context operator? For example,
  personal information about a customer should not be displayed on a
  screen in a public building.
\item \emph{Interaction invariants}: does data sensed from a given
  sensor always lead to a particular action? For example, sensing a
  fire should always cause an alarm to go off.
\end{itemize}

\subsection{Data Reachability} 

In graph theory, a vertex \texttt{y} is said to be reachable from a
vertex \texttt{x} when there is a path from \texttt{x} to \texttt{y}.
In our case, data reachability properties are of type ``A must not
access B'' or ``A may access B.'' Checking such properties is required
to ensure that, for example, private information cannot be compromised
or to identify the potential impact of a sensor failure (\eg{} a power
outage) on the rest of the application.

We define data reachability from a component in the \archipattern{} by
using the interaction contracts.

\begin{definition}[Data Reachability]
\label{def:reach}
Given a component $C$ and name $n$ of a sensor source or a
{\contextop}, the data associated with $n$ is reachable from $C$ if
one of the following conditions is satisfied:
\begin{itemize}
\margin
\item $C=n$ or
\item $C$ is a context operator and its interaction contract
  $\config{A}{U}{E}$ is such that $n$ is reachable from at least one
  of the names contained in $\msf{names}(A) \cup \msf{names}(U)$ or
\item $C$ is an actuator or a control operator and $n$ is reachable
  from one of its children.
\end{itemize}
\end{definition}

Consider an extension of the web server monitoring example where a
dedicated public web page displays the top five most visited URLs.
This top five is calculated from the data provided by the
\ct{AccessLogParser}. By applying Definition~\ref{def:reach}, we check
that the user profiles calculated by \texttt{Accessing\-Profile} are
not reachable from the actuator that updates the web page and thus
these profiles cannot be published in this public web page. When there
does exist an undesirable reachability path, the information in this
path can guide the architect in fixing the interaction contracts.

\subsection{Interaction Invariants}

Interaction invariants are properties that are verified at any state of
the \domainappl{}. For example, we would like to ensure that the
\texttt{ProfileLogger} is always activated whenever someone accesses 
the web server. We characterise the progress of an \domainappl{} by its
data flow and we use LTL~\cite{huth} (Linear Temporal Logic) formulae
to characterise interaction invariants.

For example, the property on the web server can be specified
by the following LTL formula:

$$\square (\mathit{NewLine} \rightarrow (\Diamond
\mathit{ProfileLogger\_Activated}))$$

\noindent
where the predicate \textit{NewLine} is true if a new value for the
\texttt{line} source of \texttt{AccessLogReader} is pushed and the
predicate \textit{ProfileLogger\_Activated} is true if
\texttt{ProfileLogger} is activated by a new profile.
This property can be understood as: ``\emph{At any moment}
({\scriptsize $\square$}), if a new line is pushed, then the
\texttt{ProfileLogger} will \emph{eventually} ($\Diamond$) be activated.''

To check the LTL invariants, we use the SPIN model checker associated
with the Promela modelling language~\cite{Holz03a}. If an invariant is
not satisfied, SPIN gives a counterexample in the form of an execution
trace. This counterexample can guide the architect in fixing his
interaction contracts. To check the above invariant, we translate each
interaction contract of the web server monitoring example into a
Promela process. For
example, the interaction contract
$\config{{\Uparrow(\code{AccessLogParser})}}{{\Downarrow
(\code{IP2Profile})}}{{\Uparrow \mathit{self}}}$ associated to the
\texttt{AccessingProfile} context operator is mapped into the
following Promela process specification:

{\scriptsize
\begin{verbatim}
 1: active proctype AccessingProfile() {
 2:   byte newlog, profile;
 3:   do
 4:   :: accesslogparser?newlog -> {
 5:       ip2profile_get!1; 
 6:       ip2profile_return?profile;
 7:       accessingprofile!1;
 8:     }  
 9:   od
10: }
\end{verbatim}
}

Each interaction contract is mapped into a process and each component
interaction is mapped into a channel. Also, each activation condition
is mapped into a conditional expression (line 4) which is true if
there is a new value in the channel. Each data requirement is mapped
into a sequence of two instructions: one for the transmission of the
request (line 5) and one to receive the corresponding response
(line~6). Finally the emission is translated into a message send
(line~7). The \texttt{do/od} statement is a loop construct that
encodes the reactivity of context operators.\footnote{The full Promela
  specification can be found at\\
  \url{http://diasuite.inria.fr/index.php/webserver}} The
  Promela specification is automatically generated from the
  interaction contracts. Currently, we are working on the translation
  of counterexamples given by SPIN into high-level explanations.

\subsection{Verification Support}

Data reachability and interaction invariants are two examples that show
how the architecture specification makes core concepts explicit and
thus facilitates high-level safety analyses on the data flow. These
design-time analyses support the architect by giving counterexamples.
The same properties do not have to be checked again on the
implementation because the implementation is guaranteed to conform to
the design.



\section{Evaluation}
\label{sec:discussions}

This section gives an overview of the tool suite into which we have
integrated interaction contracts, and discusses the measured benefits of
this integration.

\subsection{A Working Tool Suite}

Previously, we have developed a Domain-Specific Language (DSL),
DiaSpec, to express architecture descriptions that follow the SCC
architectural pattern~\cite{Cass09b}. This DSL is
supported by a tool suite, DiaSuite, providing code generation and
related functionalities. The code generator translates a DiaSpec
description into a Java programming framework. Applications developed
using this generated programming framework can be deployed using any
of the communication protocols supported by the available back-ends,
which currently comprise X10, UPnP, RMI and SIP. DiaSuite-compatible
drivers have been implemented for hardware such as the iPod touch,
various types of RFID tags and the Axis networked camera. Applications
can furthermore be tested prior to deployment using a 2D simulator,
requiring no change in the operator implementations. DiaSuite has been
used to develop applications in areas including home/building
automation, tier-system monitoring and
avionics.\footnote{\url{http://diasuite.inria.fr}}

Our experiences in using DiaSuite have motivated the development of
interaction contracts. Integration of interaction contracts into
DiaSuite requires extending the DiaSpec language, and correspondingly
extending the code generator to take into account the new constructs.
The resulting generated code follows the denotational semantics
presented in Section~\ref{sec:sem} and the class structure presented
in Section~\ref{sec:class_structure}

\subsection{Benefits of Interaction Contracts}

To assess the interaction contracts, we have conducted studies with 20
groups of 3 undergraduate computer science students each. The
students had no prior experience with our tool suite or SCC. We gave
each group the original version of DiaSuite~\cite{Cass09b} that did
not include the interaction contracts. We also gave all groups the
same diagram, similar to that of Figure~\ref{fig:webserver-simple}:
they had to translate it into DiaSpec and then implement the project.
All groups designed a working architecture and most completed the
assignment with an implementation. The experiment revealed some
shortcomings in DiaSpec. The rest of this section presents these
shortcomings and how interaction contracts resolve them.

\myparagraph{Design support} In the original version of DiaSpec, the
architect declared connections between components without specifying
the permitted interactions, as was illustrated with the solid arrows
in Figure~\ref{fig:webserver-simple}. In particular, the architect did
not specify whether a component is to be accessed through a push or a
pull mechanism. This imprecision lead to different interpretations of
the same architecture, and thus different implementations, some
outside of the original expectations of the architect. With the
introduction of interaction contracts, the architect precisely
expresses the allowed interactions.

\myparagraph{Programming support} We intentionally did not give
students any documentation about the generated framework, to be able
to determine to what extent this framework was in itself able to guide
the implementation. The students thus had to search in the generated
code for the methods of interest to perform component interactions.
Using the original version of DiaSpec, there are twice as many of
these methods as necessary, because the code generator is unable to
determine the intent of the architect, and thus must generate code for
both a push and a pull interaction between every pair of connected
components. Our interaction contract compiler generates abstract
methods that are self-contained: implementing them only requires using
the arguments and returning a result. Furthermore, the only abstract
methods generated are those that support the interactions intended by
the architect.

\myparagraph{Verification support} Without the interaction contracts,
the verification support is limited. We first consider reachability
properties.  Data reachability is entirely determined by the
parent-child relationship in the data-flow graph, and thus the set of
data reachable from a given component is not affected by the addition
of the interaction contracts.  Nevertheless, the introduction of
interaction contracts allows giving more precise counterexamples in
the case of reachability property violations, as the counterexample
can include the precise sequence of activation conditions. On the
other hand, most interaction invariants, such as the one shown in
Section~\ref{sec:analysis}, cannot be ensured without the interaction
contracts as there is no guarantee that a component publishes a value.

\subsection{Measuring Programming Support}

To measure the impact of interaction contracts on the degree of
programming support provided, we use several metrics on a
representative set of applications, such as the web server monitor, an
anti-intrusion system and a home remote-control application. These
applications are implemented with and without interaction contracts.
To perform these measurements, we use
Sonar,\footnote{\url{http://www.sonarsource.com/}} a platform that
uses various metrics to guide developers in improving source code
quality. As there is little variation in the measurements for the
different applications, we present only averages.

\myparagraph{Program size} For each application, we have compared both
handwritten and automatically generated number of lines of code with
the number of lines in the architecture description, the
implementation and the framework. Table~\ref{tab:effort} presents
these results. The ratio of code for the architecture description
(column \textit{Arch.}) increases slightly, because the architect must
now write the interaction contracts. The ratio of code for the
implementation (column \textit{Implem.})\ decreases, in part because
methods corresponding to useless interactions do not have to be
implemented, and in part because some functionalities, such as the
handling of synchronization (Section \ref{subsec:behavcontract}) are
moved from the implementation to the generated code, requiring less
programming effort from the application developer. Finally, the ratio
of generated code (column \textit{Framew.})\ increases slightly,
reflecting the code that has been moved from the implementation to the
generated programming framework.

\begin{table}[htbp]
  \centering
  \begin{tabular}{@{}lccc@{}}
    \toprule
    & \textbf{Arch.} & \textbf{Implem.} & \textbf{Framew.} \\
    Without interaction cont. & 6\% & 14\% & 80\% \\
    With interaction cont. & 7\% & 11\% & 82\% \\
    \bottomrule
  \end{tabular}
  \caption{The development effort for architecture descriptions without and with interaction contracts. The figures indicate the distribution (in percentage) of the number of lines of code.}
  \label{tab:effort}
\end{table}

\myparagraph{Execution coverage} Because the generated code can be
arbitrarily large without impacting development time, the measures in
Table~\ref{tab:effort} are relevant only if the generated code is
actually executed, and thus has to be produced in some manner. To
assess this, we measured the execution coverage of the programming
framework code. On average, 76\% of the generated framework is
actually executed. We studied the parts that are not executed and
found that all of them are either error handling code or features that
may not be relevant to a given application, such as entity discovery.

\myparagraph{Code quality} We also used Sonar to measure the code
quality according to various criteria including code duplication, rule
compliance, and code complexity. The results given by Sonar indicate
an overall good quality of the code written by the developer. For
example, the average code complexity of applications implemented with
the interaction contracts is 2.6, on a scale of 0 to infinity,
indicating well structured code.

These code quality results, associated with the small percentage of
code that has to be manually written, show that our generated
programming framework guides the developer in producing
well-structured and easy-to-maintain code.


\section{Related Work}
\label{sec:related}

Our work is related to software architectures, formalisms for
interaction specifications, and model-driven development.

\subsection{Software Architectures} To help in structuring
\domainappl{}s, dedicated architectural patterns have been proposed.
Chen and Klotz propose to decompose an application into information
sources and context operators~\cite{Chen02b}. Their pattern focuses on
information processing and control but does not model the relation
with the environment (\ie{} how the environment is sensed and
modified) nor does it support implementation. Architecture Description
Languages model systems to ensure various properties at compile time
and at runtime. Most ADLs are dedicated to analysing architectures and
provide little or no implementation support. Some ADLs like
Darwin~\cite{Mage96a} and Unicon~\cite{Shaw95a} generate runtime
support, but for components that have been developed separately with a
generic programming framework. These approaches provide generic
abstract classes like \ct{Component} and \ct{Connector} that the
developer must implement. As a result, component implementation cannot
benefit from any support generated from the architecture
description~\cite{Medv00a}.

In general, architecture-based approaches that support implementation
check conformance of the implementation to an architectural pattern,
\eg{} constraining the interactions between various variants of
components and connectors, but not conformance to a specific
architecture description that is an instance of that pattern. Notable
exceptions include ACOEL~\cite{Sree02a} and ArchJava~\cite{Aldr02c};
they connect ADLs and programming languages by proposing new syntactic
constructs. These constructs allow architectural concerns to be
expressed inside the application. Our work goes beyond these
approaches by separating architecture from implementation and
generating a programming framework to bridge the two.

To date, Archface~\cite{Ubay10a} is the work that is the most similar
to ours. Archface leverages concepts from Aspect-Oriented Programming
(AOP) to describe component interactions. By using
implementation-level mechanisms such as pointcuts, architects can
describe component interactions more accurately at the cost of
anticipating the structure of the implementations. As such, the
separation between architecture description and implementation becomes
blurred, making the architectural design phase more difficult. As a
general-purpose design language, Archface can be used to describe any
component-oriented architecture, which limits the support it can
provide in a particular domain.

\subsection{Interaction Specifications} Automata-based models, such
as IO automata~\cite{Lynch87} and Interface automata~\cite{Alfaro01},
are commonly used for modelling interactions and actions within
distributed and concurrent systems. These approaches have been used to
describe component interactions in ADLs~\cite{Waig08a}. Interaction
contracts are simpler in that they do not describe the full
interaction sequence but only capture interaction constraints. Our
objective is to specify only what can be enforced by the generated
framework. It is virtually impossible to completely enforce automata
behaviors via the generated framework as we cannot guess how the
developer will distribute the application logic within the sequences
of messages. We could choose to enforce it partially, but in this
case, properties verified at the automaton level would not necessarily
hold in the implementation. Moreover, automata-based models are a
general solution and do not capture the specific properties of
\domainappl{}s. For example, context operators are reactive and this
characteristic must be checked on automata, whereas it is
syntactically ensured by interaction contracts.

\subsection{Model-Driven Development} Model-Driven Development uses
models and model transformations as a way to specify software
architectures and implementations. The goal of these approaches is to
raise the level of abstraction in program specifications through
graphical notations, and to generate a working implementation from
such a specification. UML~2.0 (Unified Modeling Language) has been
widely accepted as an architecture modeling notation and as a
second-generation ADL~\cite{Medv07a}.  Some approaches, such as
PervML~\cite{Serr10a}, relies on UML diagrams and OCL expressions to
model domain-specific concerns. From such diagrams, a dedicated suite
of tools is able to generate a complete implementation of the
described system. By using UML diagrams, these approaches leverage
existing knowledge from developers and also existing tools such as the
Eclipse Graphical Modelling Framework~(GMF). Even though such
approaches propose a conceptual framework for developing applications,
they only provide the user with generic tools. The PervML approach, as
well as other MDE-based approaches, require developers to directly
manipulate OCL and UML diagrams, which become ``enormous, ambiguous
and unwieldy''~\cite{Picek08a}. In contrast, DiaSpec abstracts away
such technologies, limiting the amount of expertise required from the
developers.

The CALM framework uses models as types for component-oriented
systems~\cite{Jung10a}. CALM provides three modelling tiers, where
each tier constrains and guides activities in the tier below: the
upper tier allows the definition of domain-specific ADLs, the middle
tier allows the definition of the system components, and the lower
tier allows the instantiation and combination of these components.
DiaSpec and its notion of interaction contracts could potentially be
described in CALM's upper tier, leveraging CALM's type checking
capabilities. However, CALM neither verifies that an implementation
conforms to its architecture description nor does CALM proposes an
architect to verify safety properties.


\section{Conclusion}
\label{sec:conclusion}

In this paper, we have introduced a notion of interaction contracts
dedicated to describing SCC applications. We have shown how
interaction contracts guide the architect in describing allowed
context operator interactions. We have also described how interaction
contracts can be mapped into a generated programming framework and how
this mapping guides the implementation of SCC applications. A key
benefit of our approach is that the strategy for generating the
dedicated programming framework guarantees conformance between the
architecture and its implementation. Our generative approach allows
unlimited regeneration of the framework without overwriting the
developer's code. Finally, we have shown how interaction contracts
guide analyses at the architecture level and how the properties
checked by these analyses still hold at the implementation level.

We are currently expanding this work in several directions. We want to
further guide development by automatically generating a dedicated
unit-testing framework. Work is also in progress to add and compose
non-functional layers (\eg{} fault-tolerance, safety and security) on
top of the \archipattern{} and have automatically generated
support~\cite{Gatt11a,Merc10a}. Finally, we are investigating the
applicability of interaction contracts to other SCC component types
and to other architectural patterns.


\bibliographystyle{abbrv}
\bibliography{icse2011}  


\end{document}